\documentstyle[twoside,fleqn,espcrc2]{article}
\input epsf
\title{Properties of Low-Lying Heavy-Light Mesons\thanks{
Based on talks presented by E.~Eichten and B.~Hill}}
\author{
	Anthony Duncan,\address{Dept.~of Physics,
	Univ.~of Pittsburgh,Pittsburgh, PA~~15260~~USA}
        Estia Eichten,$^{\rm b}$ Aida X.~El-Khadra,
	\address{Fermilab, MS 106,
	PO Box 500, Batavia, IL~~60510~~USA}
	Jonathan M.~Flynn,\address{Physics Dept.,
	Univ.~of Southampton, Southampton~~SO9 5NH~~UK}
	Brian R.~Hill,\address{Dept.~of Physics,
	Univ.~of California,
	Los Angeles, CA~~90025~~USA}
	and Hank Thacker.\address{Dept. of Physics,
	Univ.~of Virginia, Charlottesville, VA~~22901~~USA}
}
\begin{document}
\begin{abstract}
We present preliminary results for $f_B$ and masses of low-lying
heavy-light mesons in the static limit.
Calculations were performed in the quenched approximation using
multistate smearing functions generated from a Hamiltonian for a
spinless relativistic quark.  The $2S$--$1S$ and
$1P$--$1S$ mass splittings are measured.
Using the $1P$--$1S$ charmonium splitting to set the overall scale,
the ground state decay constant~$f_B$, is $319 \pm 11$(stat) MeV.
\end{abstract}
\maketitle
\section{INTRODUCTION}

Lattice gauge calculations of heavy-light meson structure are of both
theoretical and phenomenological interest.\cite{Eichten}
One immediate goal of these
calculations is to obtain precise quantitative results for masses, decay
constants, and form factors in the static approximation, where the heavy
quark propagator is replaced by a timelike Wilson line. One difficulty which
plagued early, exploratory calculations of the pseudoscalar decay constant
$f_B$
was the problem of isolating the ground state contribution to the propagator
of the local weak current. Because of the proximity of excited states and
their sizeable overlap with the local current, a large separation in time was
required, with an accompanying loss of statistics.
Recent attempts to overcome this problem have employed nonlocal $\bar{Q}q$
operators (in a fixed gauge\cite{wup_fb}) smeared
over a cube\cite{rsw_fb} or wall source\cite{bls_fb}. By
measuring the asymptotic behavior of both the smeared-smeared ($SS$) and
smeared-local ($SL$) propagators, one can reduce the systematic error
associated with excited state contributions. However, it is likely that such
smearing functions are too crude to obtain accurate values of the parameters
of the low-lying heavy-light states.\cite{hs}
This is illustrated in Table~1.

\begin{table}[htb]
\centering
\caption{The overlap amplitudes between various cube smearing functions and
the approximate wavefunctions
for the ground state $|0\rangle$ and first three excited states
($|1\rangle$, $|2\rangle$,
and $|3\rangle$) for $\beta = 5.9$ and $\kappa = 0.158$ on
a $16^3$ lattice.}
\label{tab:cube}
\begin{tabular}{ccccc}
\hline
\multicolumn{1}{c}{$cube$}
&\multicolumn{1}{c}{ $|0\rangle$}
&\multicolumn{1}{c}{ $|1\rangle$}
&\multicolumn{1}{c}{ $|2\rangle$}
&\multicolumn{1}{c}{ $|3\rangle$} \\
\hline
point & 0.208 & 0.234 & 0.285 & 0.344 \\
3 & 0.568 & 0.480 & 0.362 & 0.126 \\
5 & 0.764 & 0.304 & 0.010 & -0.220 \\
7 & 0.800 & -0.024 & -0.296 & -0.197 \\
9 & 0.746 & -0.330 & -0.301 & -0.122 \\
11 & 0.663 & -0.546 & -0.099 & -0.207 \\
wall & 0.459 & -0.687 & 0.477 & 0.002 \\
\hline
\end{tabular}
\end{table}

It is important, therefore, to develop new
techniques which allow the extraction of
the properties of heavy-light states
from relatively short times.
Here we report preliminary results obtained using the multistate smearing
method discussed elsewhere~\cite{wf91,wf92}. The hallmark of a pure,
isolated ground state
meson is an effective mass plot which is constant in time. In Figs.~1
and~2, we show our results for both the $SS$ and $SL$ local effective
mass plots at
$\beta=6.1, \kappa=0.151$, on a set of (48) $24^3\times 48$ lattices.
On the horizontal axis is the time~$T$
in lattice units. On the vertical axis is $\ln(C_{SS}(T{-}1)/C_{SS}(T))$.

The $SS$~local effective
mass reaches its asymptotic value around $T=2$, while the $SL$~propagator is
nearly asymptotic after $T=3$.

The results exhibit a single consistent
plateau at $ma = 0.619 \pm 0.006$
over a large range of~$T$ for both~$SS$
and~$SL$ propagators.
The determination of the fitted value of the effective
mass will be discussed in Section~4.  We note here that the effect
of the off-diagonal entries in the covariance matrix
of the data for the two correlators has been included.\cite{Toussaint}

These mass plots convincingly demonstrate the
effectiveness of our smearing method in isolating the ground state.

\begin{figure}[htb]
\epsfbox[0 100 75 210]{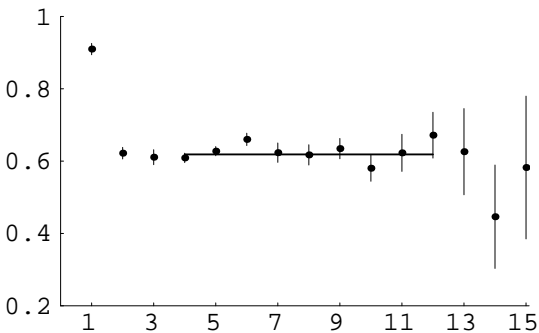}
\caption{$\kappa=0.151$ $SS$ effective mass plot.}
\label{rundee}
\end{figure}

\begin{figure}[htb]
\epsfbox[0 100 75 210]{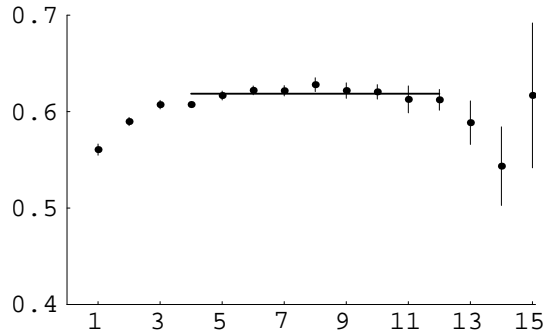}
\caption{$\kappa=0.151$ $SL$ effective mass plot.}
\label{runded}
\end{figure}

\section{WAVEFUNCTIONS}

The basics of the multistate smearing method and the choice of the
Hamiltonian to generate the smearing wavefunctions are
reported by Thacker~\cite{wf92}.  The agreement between the spinless
relativistic quark model (SRQM) wavefunctions and the lattice
QCD wavefunctions is discussed in detail there.

In the analysis presented here a two state smearing matrix was
used.  After a few iterations of the multistate smearing
method a rough value of the mass parameter $\mu$ (in the SRQM) was determined.
The output approximate ground state wavefunction at each $\kappa $
value $|0\rangle$ was extracted and used with the first excited state
$|1\rangle$
generated from the SRQM. These were the two states used as smearing functions

Two points need to be made about this variation of the multistate analysis:

(1)  The states are not exactly orthogonal.  If $x=\langle 0|1\rangle$,
then an orthonormal pair of states can be obtained by replacing $|1\rangle$ by
$|1'\rangle$ given by:
\begin{equation}
|1'\rangle = (|1\rangle - x|0\rangle )/\sqrt{1-x^2}
\end{equation}
This effect is included here but was not
included in previously presented results. The effect on
$f_B$ is approximately 12 percent.

(2) The state $|0\rangle$ varies with light quark hopping parameter $\kappa $
and this variation has some
statistical fluctuation, which is reflected in the variation with
$\kappa $ of our results.

\section{RENORMALIZATION}
In order to extract continuum results from our lattice calculation, the matrix
element calculated in the lattice effective theory must be
related to the corresponding quantity in the continuum.
This matching is done in perturbation theory in two steps.
The first step is to relate the operator (in this case the axial current)
in the continuum effective theory to its counterpart in the full theory
\cite{eh,elc}:
\begin{equation}
Z_{\rm eff} = 1 - \frac{g^2}{12\pi^2} (\frac{3}{2} \ln{\frac{m^2_b}{\mu^2}}
- 2).
\end{equation}
We use the 1992 particle data book average \cite{pdb} for
$\alpha_S \equiv g^2/4\pi$ at the scale $\mu = a^{-1}$ and mass $m_b = 5$ GeV.
In the second step the lattice current is compared to the corresponding
current in the continuum effective theory \cite{eh,elc}:
\begin{equation}
Z_{\rm lat} = 1 + \frac{g^2}{12\pi^2} 20.37.
\end{equation}
The renormalization of the axial current is then $Z_A^{-1}=Z_{\rm eff}Z_{\rm
lat}$.
Table~\ref{tab:renorm} lists the renormalization constants for all the lattices
analyzed here. We use the results for the 1P--1S splitting in charmonium
to obtain $a^{-1}$ \cite{prl}.

\begin{table}[htb]
\centering
\caption{The current renormalizations for different lattices}
\label{tab:renorm}
\begin{tabular}{cccccc} \hline
{\rm lattice} & $\beta$ & $Z_{\rm eff}$ & $Z_{\rm lat}$ & $Z_A$ \\
\hline
$12^3\times 24$ & 5.7 & 0.90 & 1.33 & 0.83 \\
$16^3\times 32$ & 5.9 & 0.96 & 1.31 & 0.79 \\
$24^3\times 48$ & 6.1 & 1.00 & 1.30 & 0.77 \\ \hline
\end{tabular}
\end{table}
\section{ANALYSIS}

We now discuss the statistical analysis leading to our results
for~$f_B$.
In the first and second subsections below,
we will explain the methodology leading to the central values
and statistical errors quoted in our abstract and conclusions.
In the third subsection, we will look at variations in the analysis
procedure and their impact on these results. Variations are
partly attributable
to the remaining systematic errors in the computation, allowing
an estimate of the magnitude of these errors.

We present results from three different lattices as listed
in table~\ref{tab:lat}.  The gauge configurations are separated
by 500 ($12^3\times 24, \beta=5.7$), 2000 ($16^3\times 32, \beta=5.9$),
and 4000 ($24^3\times 48, \beta=6.1$) pseudo-heat-bath sweeps respectively.
They were fixed to machine accuracy in Coulomb
gauge using Fourier acceleration.
We use the Wilson action for the light quarks.
We present results from
a single $\kappa$ value at $\beta=5.7$ and $\beta=6.1$.  At
$\beta=5.9$, the range $\kappa=0.154-0.159$ includes~0.154, 0.156, 0.157,
0.158, and~0.159.

\begin{table}[htb]
\centering
\caption{The lattices and parameters}
\label{tab:lat}
\begin{tabular}{cccl} \hline
{\rm lattice} & $\beta$ & {\rm confs.} & $\kappa_{\rm light}$ \\
\hline
$12^3\times 24$ & 5.7 & 48 & 0.165 \\
$16^3\times 32$ & 5.9 & 48 & 0.154--0.159 \\
$24^3\times 48$ & 6.1 & 48 & 0.151 \\
\hline
\end{tabular}
\end{table}
\subsection{Effective Masses and Selection of Fit Range}

The effective mass plots for the $24^3\times 48$ lattices shown in Figs.~1
and~2 exhibit the asymptotic ground state signal starting at quite
small times.
However, at small times, even with carefully chosen wave functions, there are
significant contributions to smeared source--smeared sink ($SS$) and smeared
source--local sink ($SL$) correlators
from higher energy states with the same quantum numbers as the $B$ meson.
This is apparent below time $3$ in the $\beta=5.9$ $\kappa=0.158$ $SS$
effective mass plot,
\begin{figure}[htb]
\epsfbox[0 100 75 210]{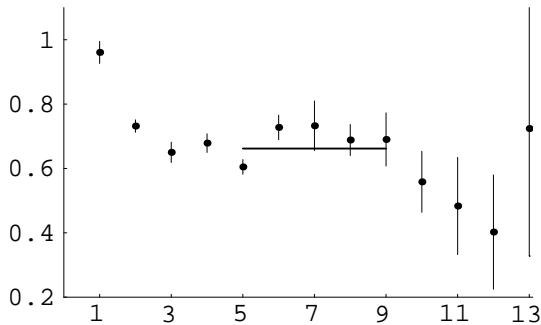}
\caption{$\kappa=0.158$ $SS$ effective mass plot.}
\label{runcee}
\end{figure}
depicted in Figure~\ref{runcee}.

{}From time slice~3 on, the local effective mass is consistent
with a constant function of time.  This is plausible from
inspection of the plot itself, which has the rms (not jackknife) errors
plotted in addition to the means.  To confirm this, we
fit the logarithm of the $SS$ correlator
to a linear function over the range of times 2--9, and found
$\chi^2$=8.9 for 6 degrees of freedom.
If the lower limit of the fit range is reduced to~1, $\chi^2$ increases
to~12.1 for~7 degrees of freedom.
These values of $\chi^2$ are obtained from
a single fit using 48 gauge field configurations.
To obtain the effective mass and its statistical error,
this fit is jackknifed using~12
subensembles, each with~4 of the~48 lattices removed.
Using the fit range 2--9, we find $ma$=0.664 $\pm$ 0.011 (statistical).

We now turn to the corresponding $SL$ effective mass plot,
\begin{figure}[htb]
\epsfbox[0 100 75 210]{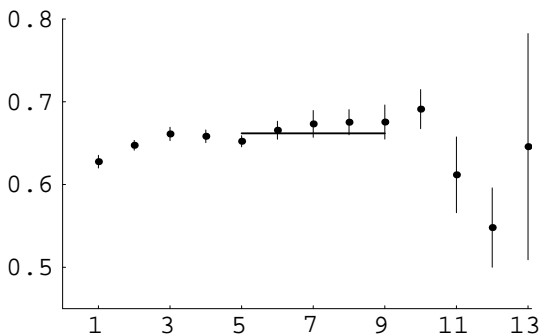}
\caption{$\kappa=0.158$ $SL$ effective mass plot.}
\label{runced}
\end{figure}
depicted in Figure~\ref{runced}.  This correlator has
much smaller fluctuations (note the change in scale for the ordinate),
and is capable of revealing
statistically significant variations in the
local effective mass over the range 2--9.
Such variations are apparent in effective mass
plots at lower values of~$\kappa$ (not depicted),
and can be ascribed
to an admixture of higher energy states with the same quantum numbers as the
$B$~meson.
Calculations of
$\chi^2$ for various fits at $\kappa=0.154$ suggest that the lower
limit of the
fit range should be increased to~5.
Using the fit range 5--9, for the $\kappa$=0.158 $SL$ correlator we find that
$ma=0.667$ $\pm$0.009 (jackknife).

The $\kappa=0.158$ effective masses from the two correlators
appear to be
consistent to within one standard deviation.  However, monitoring $\chi^2$
for a simultaneous fit that requires the two correlators to have a common
effective mass
gives a more precise indication, since the statistical fluctuations of
the $SS$ and $SL$ correlators may be correlated.  We therefore
performed fits on all $16^3\times 32$ data (five $\kappa$ values)
using a large variety of fit intervals (demanding a common slope for
the corresponding $SS$ and $SL$ correlators).
We looked for fit
intervals of at least three units of time which
had $\chi^2$ per degree of freedom near one.
We selected the lowest value of
the lower limit of the fit range that met these conditions.
For the $16^3\times32$ data, the fit range so selected was 5--9.
For the single $\kappa$ value at $12^3\times24$, we selected 3--9,
and for $24^3\times48$, we selected 4--12.
Intra-kappa correlations were
included\cite{Toussaint}.
We will discuss the effect of variations of the fit range
and inter-kappa correlations in Subsection~4.3.
\subsection{Results for $f_B$}
In Figures~\ref{runcee} and~\ref{runced}
the horizontal line superimposed
on the local effective masses is the common mass obtained from a
simultaneous jackknife
fit to the $SS$ and $SL$ correlators using the fit interval 5--9.  The
$\kappa$=0.158 results are representative of the results for each of the
five $\kappa$ values.
\begin{table}[hbt]
\setlength{\tabcolsep}{0.58pc}
\newlength{\digitwidth} \settowidth{\digitwidth}{\rm 0}
\catcode`?=\active \def?{\kern\digitwidth}
\caption{Fitted parameters as a function of $\kappa$ at $\beta = 5.9$}
\label{tab:fbma}
\begin{tabular}{lrrrrr}
\hline
$\kappa$            & 0.159 & 0.158 & 0.157 & 0.156 & 0.154  \\
\hline
$f_B$               &  324  &  316  &  356  &  344  &  364   \\
                    & $?11$ & $?10$ & $?11$ & $?10$ & $?10$  \\
$ma$                & 0.659 & 0.662 & 0.687 & 0.687 & 0.710  \\
                    & 0.014 & 0.010 & 0.009 & 0.008 & 0.007  \\
$m'a$               & 0.940 & 0.935 & 0.950 & 0.945 & 0.954  \\
                    & 0.015 & 0.013 & 0.013 & 0.012 & 0.012  \\
\hline
\end{tabular}
\end{table}
Once the common mass and two intercepts have been obtained, the value
of $f_B$ at each $\kappa$ value is determined from the intercepts
of the $SS$ and $SL$ fits.  These results are presented in
Table~\ref{tab:fbma}, which also contains the effective mass of the
meson in lattice units as a function of $\kappa$.  Below each quantity
is its jackknife uncertainty. $m'$ will be discussed in Section~5.

We then extrapolate $f_B$ to $\kappa_c$
the critical value of $\kappa$.
We measured $\kappa_c=0.1597 \pm 0.0001$.
Since the dependence on $\kappa$ is weak, the uncertainty
in $\kappa_c$ has a negligible (of order 1~MeV) effect on the results.
\begin{figure}[htb]
\epsfbox[0 100 75 210]{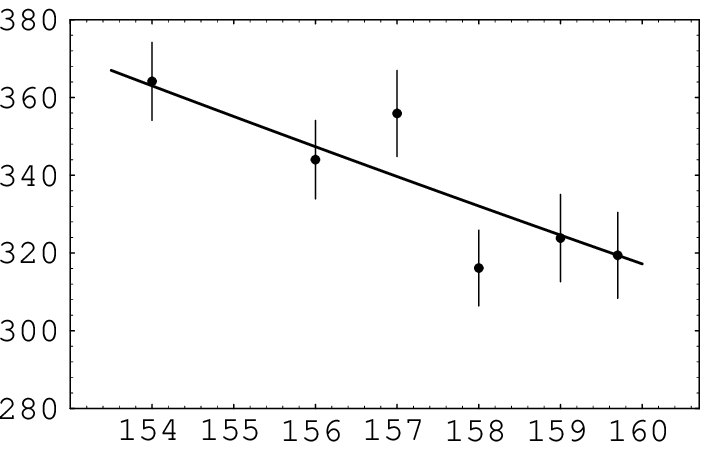}
\caption{$f_B$ as a function of $\kappa$.}
\label{five}
\end{figure}
The results for the extrapolation as well as the results in
Table~\ref{tab:fbma}\
are plotted in Figure~\ref{five}.
At $\kappa_c$, we find for the decay constant
$f_B =  319 \pm 11~{\rm MeV~(jackknife).}$
The slope with respect to $\kappa^{-1}$
is 188 $\pm$ 32~MeV.  The extrapolated value of $f_B$ and slope with respect
to $\kappa^{-1}$ are superimposed on the Monte Carlo data.  The value plotted
at $\kappa$=0.1597 is the extrapolated value of $f_B$ with its errors.
The nonlinearity from using $\kappa$ as the
horizontal axis rather than $\kappa^{-1}$ is imperceptible in the
superimposed fit, and were we to have fit linearly in $\kappa$ rather than
$\kappa^{-1}$ we would have changed the extrapolated value of $f_B$ and the
(transformed) slope negligibly relative to the statistical errors.
We note that the $\chi^2$ was 51 for 44 degrees
of freedom.

We conclude this section by noting our results for
$f_B$ in the static limit from $\beta=5.7$ $\kappa=0.165$
and $\beta=6.1$ $\kappa=0.151$.  In the former case using the fit
range 3--9, we find $f_B$=351 $\pm$ 13 MeV
($a^{-1}$= 1.15~GeV)
and in the latter case using the fit range 4--12, we find $f_B$=359 $\pm$ 7 MeV
($a^{-1}$= 2.43~GeV).
Using the pion mass to determine the corresponding value
of $\kappa$ for $\beta = 5.9$ we obtain from our fit to the
$\beta = 5.9$ data the corresponding values of $f_B$.  They are
$f_B = 330$ MeV for $\beta = 5.7$ and $f_B = 367$ MeV
for $\beta = 6.1$.
\subsection{Dependence on Analysis Procedure}
In this subsection, we investigate the dependence of our results
on (1) the fit interval, and (2)
the inclusion of inter-kappa entries in the covariance matrix.  Perhaps
the most interesting of the variations in analysis is changing the
admixture of the excited state.  This will be discussed in the Section 5.

The dependence of $f_B$ on fit interval has been investigated by rerunning
the analysis over the other fit ranges, in addition to the
primary range~5--9.  We have reproduced the analog of
Figure~\ref{five}\ in
\begin{figure}[htb]
\epsfbox[0 100 75 210]{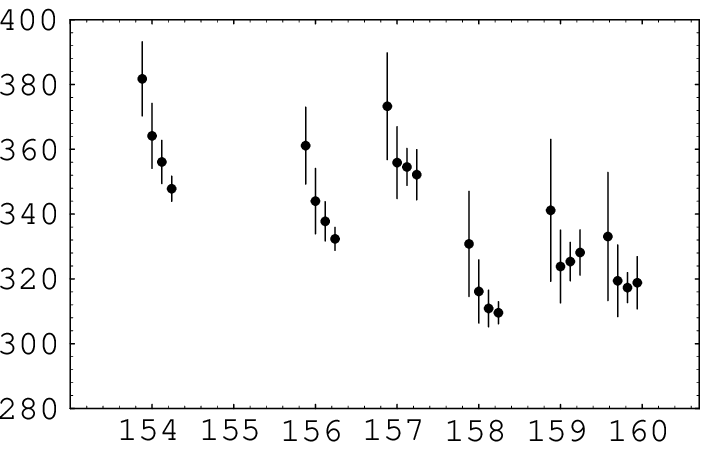}
\caption{$f_B$ as a function of $\kappa$.}
\label{intervalthree}
\end{figure}
Figure~\ref{intervalthree}.  The results from the additional fit
ranges are displaced slightly from their
true $\kappa$ values.  From left to right, they are 6--10, 5--9, 4--8,
and~3--7.
In general, the values are consistent with those of the primary fit
range, 5--9, except at small values of $\kappa$ where it has
already been noted that the $\chi^2$ per degree of freedom indicates
that excited states are contributing when the lower limit of the fit
range is less than~5.

We now consider the impact of inter-kappa
correlations.
It is apparent from comparing the $\kappa=0.158$ effective
mass plots with those of other $\kappa$ values (not depicted),
that there are
strong correlations between corresponding quantities at different $\kappa$
values.  The reason these were not included in our analysis is that
with 50 data points per configuration (5 $\kappa $'s and 5 T's for SS and SL)
it is impossible to compute a nonsingular covariance matrix.
To investigate the effect on the extrapolated value of $f_B$, we therefore
reduced the number of $\kappa$ values to~2, selecting $\kappa$=0.158 and
$\kappa$=0.156.  We then performed the fit with and without the inter-kappa
entries of the covariance matrix, and found that a
decrease in the fitted and extrapolated values of $f_B$ resulted from including
inter-kappa correlations. However these effects are in all cases
less than the one sigma level.
Larger statistics would allow us to study the impact of
inter-kappa correlation with more $\kappa$ values in the simultaneous fit.
\section{EXCITED STATES}
In this section, we will examine evidence for
the first radially excited state of the $B$
meson, and study the impact of the admixture of this wave function.
Preliminary results for the orbitally
excited states are presented in the second subsection.
Finally, finite volume systematics are discussed in the last subsection.

\subsection{Radial Excitations}

We begin this subsection by looking at the effect on the fitted values
of $f_B$ of the admixture of
the states created by the first radially excited state smearing function.
The smearing function
for the analyses in the preceding section was the linear
combination of smearing functions which diagonalized a two-by-two
matrix of $SS$ correlators which was averaged over the same range
as the primary fit range.  For the $16^3\times32$ data, the admixture of
the trial first excited state increased
from 5\% to 7\% with increasing $\kappa$.

To see the impact of this 5 to 7\% admixture on the final
\begin{figure}[htb]
\epsfbox[0 100 75 210]{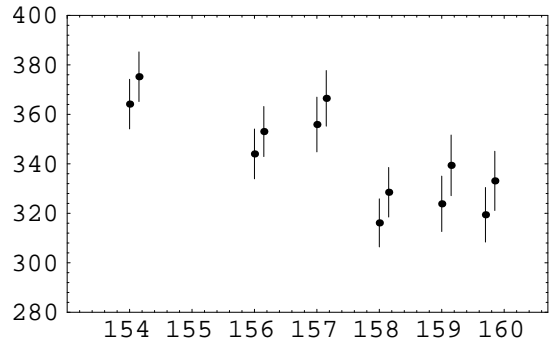}
\caption{$f_B$ as a function of $\kappa$.}
\label{undiag}
\end{figure}
results, in Figure~\ref{undiag} we have replotted the data shown in
Figure~\ref{five}\ along with the values of $f_B$ obtained from
the undiagonalized smearing function.  The additional data
is displaced slightly from its correct $\kappa$ value for visibility.
The extrapolated value of $f_B$ using the undiagonalized smearing function
is 333 $\pm$ 12~MeV, a 4\% increase.

We now examine the linear combination of the trial ground and
first excited states which is orthogonal to that used for the ground state.
Since we do not have correlators involving states with higher radial
excitations, we expect that this orthogonal state is missing contributions
from higher radial excitations of the same order as the mixing of the trial
ground and first excited state. It is neverthelesss interesting
\begin{figure}[htb]
\epsfbox[0 100 75 210]{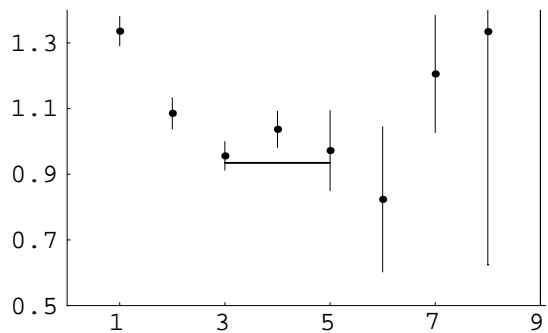}
\caption{$\kappa=0.158$ $SS$ excited state
effective mass.}
\label{excitee}
\end{figure}
to study the effective mass plots
obtained from this orthogonal linear combination, which is mostly
the trial first excited state.
In Figure~\ref{excitee}\ we plot the $SS$ local effective mass for
\begin{figure}[htb]
\epsfbox[0 100 75 210]{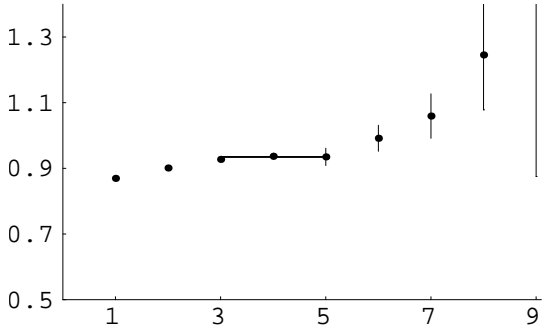}
\caption{$\kappa=0.158$ $SL$ excited state
effective mass.}
\label{excited}
\end{figure}
this correlator at $\kappa=0.158$, and in Figure~\ref{excited}\ we
plot the corresponding $SL$ correlator.
The same criteria to determine the fit interval as in the ground
state were used, and gave a fit interval from 3--5.
The excited state effective masses in lattice units, $m'a$,
using this fit range are presented in Table~4.
As usual, this is a common effective mass fitted to both the $SS$ and $SL$
correlator.  The $2S$--$1S$ splitting in lattice units,
$m'a{-}ma$, decreases from 0.281 to 0.244 as $\kappa$ goes from
0.159 to 0.154.

The decay constants of the excited state for the five $\kappa$ values
are approximately 600~MeV
with a variation in $\kappa$ and jackknife errors both less than
$\pm 15$~MeV.
A slight trend is toward larger decay constants with
increasing $\kappa$.
The value of the decay constant, $f_B'$, extrapolated to $\kappa_c$
is $618\pm11$~MeV~(stat).
The quality of the plateau is much less convincing than for the
$\beta=5.9$ ground state.
The two-state approximation may induce larger
systematic errors here than for the ground state.
In addition, finite volume effects
may be larger for the excited states.
One indication supporting these preliminary results is that the
values of the effective mass and decay constant are increased by
about one sigma only if the fit range is changed to 4--6.

\subsection{Orbital Excitations}

The lowest P wave heavy-light mesons
have light quark total angular momenta
$j_l = 1/2$ and~$3/2$. Each of these states is
degenerate with the $J = 0,1$ for the $j_l = 1/2$
state and $J = 1,2$ for the $j_l = 3/2$ state.
Smearing functions for these P wave heavy-light mesons
can be generated from the SRQM with the mass parameter $\mu $
which gives the best fit for the S wave ground state.

To date only a preliminary analysis has been performed on the
P waves. The results for the mass splittings are given in Table~5.
\begin{table}[htb]
\centering
\caption{The mass splittings (in MeV) for the lowest heavy-light P states
at $\beta = 5.9$ for various light quark $\kappa $ values.}
\begin{tabular}{c|c|c}
\hline
\multicolumn{1}{c|}{$\kappa$}
&\multicolumn{1}{c|}{$1P_{1/2}-1S$}
&\multicolumn{1}{c}{$1P_{3/2}-1P_{1/2}$} \\
\hline
0.158 & $386\pm 90$ & $43\pm 102$ \\
0.156 & $395\pm 86$ & $59\pm 95$ \\
0.154 & $416\pm 81$ & $62\pm 90$ \\
\hline
\end{tabular}
\end{table}
Considerable refinement will be required before the
splitting between the $j_l = 1/2$ and $3/2$ states can be
observed clearly.

\subsection{Finite Volume Corrections}

The systematic effects of finite volume are under study.  It would
be expected that these effects are more pronounced for the excited
states than for the ground state because the RMS radius for excited
states is larger than for the ground state.  The excellent
agreement between the SRQM wavefunctions and the measured wavefunctions
of the $1S$, $1P$, and $2S$ heavy-light states reported by Thacker~\cite{wf92}.
allows us to estimate the effects of finite volume with our periodic
boundary conditions using SRQM results.  We calculated the static energies
using 48 Coulomb gauge fixed configurations at $\beta = 5.9$
for lattices of spatial size $12^3$, $16^3$, and $20^3$.  Using
the mass parameter $\mu $ needed in the SRQM (obtained
from our study of the heavy-light mesons on the $16^3\times 32$
lattices)
we estimated the effects on various mass and wavefunction parameters at
the other two volumes.  The results are shown in Table~6.
\begin{table}
\centering
\caption{Finite volume effects in the SRQM. Variation of
eigenvalues and properties of the eigenstates for the ground and
radially excited states. Results are at $\beta = 5.9$. $\epsilon_n$
denotes the eigenvalue for the $1S (n=0)$ and $2S (n=1)$ states.}
\begin{tabular}{cccc}
\hline
\multicolumn{1}{c}{Measure}
&\multicolumn{1}{c}{$12^3$}
&\multicolumn{1}{c}{$16^3$}
&\multicolumn{1}{c}{$20^3$}\\
\hline
$\epsilon _0$ & $0.900$ & $0.912$ & $0.911$ \\
$\epsilon _1 - \epsilon _0$ & $0.251$ & $0.278$ & $0.277$ \\
$|\Psi_0(0)|$ & $0.193$ & $0.209$ & $0.206$ \\
$|\Psi_1(0)|$ & $0.239$ & $0.236$ & $0.236$ \\
$<r_0^2>^{1/2}$ & $3.320$ & $3.01$ & $3.03$ \\
$<r_1^2>^{1/2}$ & $5.09$ & $5.50$ & $5.53$ \\
\hline
\end{tabular}
\end{table}
The typical
variation between results at $12^3$ and $16^3$ are 10\% or more,
while the variation between $16^3$ and $20^3$ have dropped to a few
percent.  The validity of these model results is presently being
checked by a complete analysis of heavy-light mesons on the
$12^3\times 24$ and $20^3\times 40$ lattices.

\section{CONCLUSIONS}

By using good approximate wave functions in a multistate smearing
calculation, it is possible to control the systematic errors associated
with extracting the decay constant from nonasymptotic time. Our results
are very encouraging, not only for the present calculations, but for the
determination of other B-meson decay parameters and form factors. All such
calculations require the external meson to be in a pure eigenstate (i.e.
on shell). By using the optimized smearing functions discussed here, this
requirement can be met with a minimum separation between the B-meson source
and the electroweak vertex, greatly improving the precision of the results.

We find that at $\beta = 5.9$ and $\kappa =
\kappa_{\rm critical}$
\begin{equation}
   f_B = 319 \pm 11({\rm stat})\times {Z_A \over 0.79}\times
    \left ({a^{-1} \over
      1.75 {\rm GeV}} \right )^{3/2} {\rm \hspace{-.4cm}MeV} \;
\end{equation}
The efficacy of our smearing method is even greater at $\beta = 6.1$ as can be
seen by comparing Figs.~1 and~2 with Figs.~3 and~4 with the same
physical volume.

In addition to ground state properties, this method allows
extraction of the properties of
the lowest radially and orbitally excited states.
Our results for these
quantities are preliminary, and we expect more precise estimates to be
obtained by further analysis.

Further numerical studies are required to check the finite volume
effects, determine the effect of using an improved action for the light
quarks, and include additional light quark mass values at
$\beta = 6.1 $ which will allow a better determination of the
scaling behaviour of these properties of low-lying heavy-light
states.

{\noindent\bf ACKNOWLEDGEMENTS}

We thank George Hockney, Andreas Kronfeld, and Paul Mackenzie
for joint lattice efforts without which this analysis would not
have been possible.  JMF thanks the Nuffield Foundation for support under
the scheme of Awards for Newly Appointed Science Lecturers.
The numerical calculations were performed on the Fermilab
ACPMAPS computer system developed by the CR\&D department in collaboration
with the theory group. This work is supported in part by
the Department of Energy under Contract Nos.~DE--AT03--88ER 40383
Mod A006--Task C and DE-AS05-89ER40518, and the National Science Foundation
under Grant No.~PHY-90-24764.

\end{document}